\documentclass[intlimits,twoside,a4paper]{article}
\usepackage[cp1251]{inputenc}

\usepackage[eqsecnum]{cmpj3}

\usepackage{bm}

\issue{2019}{22}{2}{23704}
\doinumber{10.5488/CMP.22.23704}

\title[Size effects of a nanoobject in magnetic field]{Size effects of a nanoobject in magnetic field}
\author{B.A. Lukiyanets, D.V. Matulka}
\address{Lviv Polytechnic National University, 12 S. Bandera St., 79000 Lviv, Ukraine
}

\date{Received March 19, 2019, in final form May 2, 2019}

\DeclareMathOperator{\rot}{rot}

\begin{document}

\maketitle

\begin{abstract}
A theoretical analysis of physical properties of the effect of size of a nanoobject in the form of a rectangular parallelepiped whose sides $a$, $b$, $c$, are oriented along the $OX$, $OY$, $OZ$, respectively, is carried out. In the framework of the perturbation theory, changes in the electronic spectrum of the nanoobject caused by an external magnetic field $\vec{B}$, depending on its size, are analyzed. We consider two cases of the fields which are described 1) by the Landau gauge, $\vec{A}(\vec{r})=\left(0,Bx,0\right)$ ($\vec{B}$ is oriented along the side $c$) and 2) by  $\vec{A}(\vec{r})=\left(Bz,0,\alpha By\right)$ ($\alpha$ is a parameter; at  $\alpha  = 0$, $\vec{B}$ is directed along $OX$ axis, and  at $\alpha = 1$, $\vec{B}$ is directed along the diagonal in $XOY$ plane). Firstly, it is shown that the first correction to the spectrum is zero, regardless of $\vec{B}$ orientation. Secondly,  it is established that, in contrast to the case of the field orientation 1), where the correction does not depend on the length of $c$, in the case 2) such  correction  depends both on $c$ and on its ratios to the lengths of $a$ and $b$. There  was found the existence of such nanoobject sizes in $XOY$ plane at which the corrections to the spectrum are the same for different lengths of $c$ of the nanoobject.
\pacs 73.22.-f

\end{abstract}

\section{Introduction}

The influence of the shape and size of nanoobjects on their physical properties (electronic, optical, optoelectric, and others) is an important problem in modern nanotechnology.

The electronic spectrum of crystals is the basis of microelectronics. It is known that the   spectrum  of an infinite crystal is quasi-continuous, while the   spectrum  of a nanosized crystal  is noticeably discrete. Such a quantized character can be practically used in microelectronics. Moreover, if the size of a nanoobject is equal to the length of the coherence of electronic excitations then, by changing its size and/or shape, we can manipulate with its basic characteristics. Such changes in combination with optoelectric characteristics create the basis of such a promising scientific and technological direction as femtosecond optics \cite{Sto14}.

There are two methods of fabrication of nanoobjects in modern nanotechnology \cite{Pra07}: 
1) top-down nanotechnology --- the transition from bulk crystal to nanosized one by removing its layers;
2) bottom-up nanotechnology --- fabrication of a nanoobject at an atomic level by addition of atom to atom, molecule to molecule or cluster to cluster. Until the 90s of the last century, semiconductor quantum dots (zero-dimensional objects) were fabricated as dopants in glasses and in solvates or as epitaxially grown objects. While in the former case the quantum dots were of a nearly spherical shape with a diameter of 1\ldots10~nm, epitaxial quantum dots were of an elongated shape of the order of nm and a cross-section of about 10~nm. However, such nanoobjects had a wide-size spread, which limited the accuracy of their experimental studies. After 90s,  high-quality, practically monodisperse quantum dots were achieved by injection of highly reactive organometallic precursors into solvent.

Modern technology allows us to change their size and shape, and thus, purposefully change their optical and photoelectric characteristics, cathodoluminescence, Raman scattering, thermodynamics. An analysis of such effects occupies a significant place in the monographs \cite{Yam11,Cot11,Kob11,Iss11,Ant11,Mor11,Kla11}. Cognition of the physics of sizes and shapes in nanoobjects remains urgent.  In \cite{Sin18}, the theoretical investigation of changes in the band gap of the semiconductor nanocrystals XY (X --- Cd, Zn, Y --- S, Se, Te)  related to their shape and sizes is presented. Similar changes are recorded in \cite{Abd12} in the study of GaP nanocrystals. In \cite{Suk14}, electron and hole states of InAs nanocrystals of the same size, but of different ellipsoidal shape, were analyzed. It is shown that the changes in the energy states lead to significant differences in the spectra of radiation with different polarizations. Results \cite{Lif14} showed differences in the optical and electrical characteristics of the colloidal semiconductor nanocrystals fabricated in the form of spherical quantum dots, nanowires, two-dimensional nanoplates, and nanosheets. In \cite{Hel16}, it was found that silver nanoparticles of different shapes (sphere, cube, rod), but of the same size, have different biological characteristics.  An analysis of the differences between the electronic spectra of quantum dots and quantum wells is given in \cite{LiJ09}. Recent works \cite{Hol17, Hol18} present the results of investigations of the influence of a magnetic field on the electronic spectrum and on optical properties of nanoobjects. Even such an incomplete list of publications on the effects of shapes and sizes in nano-sized materials is a convincing proof of its relevance.

Herein below we will analyze, within the perturbation theory, the change of the electronic spectrum of a nanoobject due to an applied magnetic field, depending on the size of the nanoobject.

\section{Model. Calculations}
Let us consider a nanoobject of the shape of a rectangular parallelepiped with the sides $a$, $b$, $c$ located along  $OX$, $OY$, $OZ$ axes, respectively. Let the potential energy of an electron in it be the sum of three terms
\begin{align}
\label{1}
U\left(\vec{r}\right)=U\left(x\right)+U\left(y\right)+U\left(z\right),
\end{align}
where each of them is an infinite deep square well, i.e.,
\begin{center}
$$
U(s)=\begin{cases}
0,&\text{if $s=x \in [0,a]$ or $y \in[0,b] $ or $z \in [0,c]$},\\
\infty,&\text {otherwise.}
\end{cases}
$$
\end{center}

In this case, the electronic spectrum  is a solution of the time-independent Schr\"odinger equation with the Hamiltonian

\begin{align}
\label{2}
\Hat{H}_{0} =\frac{\vec{p}^{\,2}}{2m}+U(\vec{r})
\end{align}
and equals \cite{Dav76}
\begin{align}
\label{3}
E(n_{a} n_{b} n_{c})=\sum _{s=a,b,c}\frac{\hbar ^{2} \piup ^{2} }{2m}  \frac{n_{s}^{2} }{s^{2} }\Rightarrow \frac{\hbar ^{2} \piup ^{2} }{2m} \left(\frac{n_{a}^{2} }{a^{2} } +\frac{n_{b}^{2} }{b^{2} } +\frac{n_{c}^{2} }{c^{2} }\right),
\end{align}
where $n_{s} =1, 2, 3,\dots$ are principal quantum numbers. It is seen that the spectrum (\ref{3}) is a set of discrete energy levels separated by a distance, dependent on the widths of the wells. Moreover, at $a=b=c$, each state is triply degenerate, at $a=b\ne c$, $a=c\ne b$ or $b=c\ne a$ each state is doublet plus singlet level, whereas at $a\ne b\ne c$, each state is a set of three singlet levels.

The wave functions of an electron in such states are as follows:
\begin{align}
\label{4}
\psi \left(xyz\right)=\sqrt{\frac{2}{a} } \sin \frac{n_{a} \piup }{a} x\cdot \sqrt{\frac{2}{b} } \sin \frac{n_{b} \piup }{b} y\cdot \sqrt{\frac{2}{c} } \sin \frac{n_{c} \piup }{c} z.
\end{align}

 Let us place the nanoobject in a uniform magnetic field  whose vector potential is $\vec{A}\left(\vec{r}\right)$.
 It is known \cite{Dav76, Gru06} that, in the general case, the Hamiltonian of the electronic subsystem in the magnetic field $\vec{B}$ takes the following form:
\begin{align}
\label{5}
\hat{H}=\frac{\big[\vec{p}+e\vec{A}\left(\vec{r}\right)\big]^{2} }{2m} +U\left(\vec{r}\right)+\left[\frac{e\hbar }{2m} \vec{B}+\vec{\alpha }\left(\vec{r}\right)\right]\vec{\sigma }.
\end{align}
Here, $\vec\sigma=\left(\sigma _{x} \sigma _{y} \sigma _{z} \right)$ where $\sigma _{x} =\left(\begin{array}{cc} {0} & {1} \\ {1} & {0} \end{array}\right)$, $\sigma _{y} =\left(\begin{array}{cc} {0} & {-i} \\ {i} & {0} \end{array}\right)$, $\sigma _{z} =\left(\begin{array}{cc} {1} & {0} \\ {0} & {-1} \end{array}\right)$  are Pauli matrix; and  the expression $\vec{\alpha}\left(\vec{r}\right)=\frac{\hbar }{4m^{2}} \{\vec{\nabla }U\left(\vec{r}\right)\cdot [\Hat{\vec{p}}+e\vec{A}(\vec{r})]\}\vec{\sigma}$ describes spin-orbital interaction.

After transformations, the Hamiltonian (\ref{5}) can be rewritten as follows:
\begin{align}
\label{6}
\hat{H}=\hat{H}_{0} +\hat{H}_\text{p}\,,
\end{align}
where
\begin{align}
\label{7}
\hat{H}_\text{p} =\frac{e}{2m} \left[\vec{p}\vec{A}\left(\vec{r}\right)+\vec{A}\left(\vec{r}\right)\vec{p}\right]+\frac{e^{2} }{2m} A^{2} \left(\vec{r}\right)+\frac{e\hbar }{2m} \vec{B}\vec{\sigma }+\vec{\alpha }\vec{\sigma}
\end{align}
is the perturbation Hamiltonian which describes the interaction of an electron with the magnetic field.

The second term in (\ref{7}) can be neglected in comparison with the first one. The last term in (\ref{7}), taking into account the potential (\ref{1}), is absent, because $\vec{\nabla}U\left(\vec{r}\right)=0$. Therefore,
\begin{align} \label{8}
\hat{H}_\text{p} =\frac{e}{2m}\left[\vec{p}\vec{A}\left(\vec{r}\right)+\vec{A}\left(\vec{r}\right)\vec{p}\right]+\frac{e\hbar }{2m} \vec{B}\vec{\sigma }.
\end{align}
The perturbation theory criterion, i.e., ${\left\langle f \right|} \hat{H}_{1} {|i\rangle}\ll|E_{f}-E_{i} |$, in nanoobjects with their sharp discrete spectrum, permits consideration of a problem with magnetic fields in a wide range of their values. Then, according to \cite{Gru06},
\begin{align}
\label{9}
{\left\langle n_{a} n_{b} n_{c}\right|}\hat{H}{\left| n_{a} n_{b} n_{c}  \right\rangle} &=\Big\langle n_{a} n_{b} n_{c} \Big| \hat{H}_{0}+\frac{e\hbar}{2m}\vec{B}\vec{\sigma }\Big| n_{a} n_{b} n_{c} \Big\rangle +{\left\langle n_{a} n_{b} n_{c}  \right|} \hat{H}_{1} {\left| n_{a} n_{b} n_{c} \right\rangle}
\nonumber\\
&+\sideset{}{’}\sum_{n_{x} n_{y} n_{z}}\frac{\langle n_{a} n_{b} n_{c}| \hat{H}_{1} | n_{x} n_{y} n_{z}\rangle \langle n_{x} n_{y} n_{z}| \hat{H}_{1}|n_{a} n_{b} n_{c} \rangle}{E_{n_{a}n_{b}n_{c}}-E_{n_{x} n_{y}n_{z}}}\,,
\end{align}
where $\hat{H}_{1}=\hat{H}_\text{p}-\frac{e\hbar}{2m}\vec{B}\vec{\sigma }$. The 1st term is the sum of the matrix elements $\langle n |\hat{H}_{0}| n\rangle$ and  $\langle n |\frac{e\hbar }{2m} \vec{B}\vec{\sigma }| n\rangle $  (here, $\left|n \right\rangle = \left|n_{a}n_{b}n_{c}\right\rangle \left|s \right\rangle$; $\left|s \right\rangle = \left|\uparrow \right\rangle$ or  $\left|\downarrow \right\rangle$ are spin indices), and the last two ones describe the perturbation up to the second order correction. In the 2nd correction, summation is carried out in all states, with the exception of  $n_{x}n_{y}n_{z}=n_{a}n_{b}n_{c}$.
The above-mentioned matrix element $ \langle n |\frac{e\hbar }{2m} \vec{B}\vec{\sigma }| n\rangle $  in Landau  gauge case, due to the absence of spin-orbital interaction after summation over spin indices, takes the form: 
$\langle n |\frac{e\hbar }{2m} \vec{B}\vec{\sigma }| n\rangle \Rightarrow \langle n |\frac{e\hbar}{2m} B_{z} \sigma _{z} | n\rangle \Rightarrow \pm \frac{e\hbar}{2m} B_{z} \langle n {| n \rangle} \Rightarrow \pm \frac{e\hbar }{2m} B_{z} $. The obtained result does not depend on the size of the nanoobject and  it is a value of the displacement of any state $E( n_{1}n_{2}n_{3})$   (\ref{3}) up and down with the spin  $\left|\uparrow \right\rangle$  and $\left|\downarrow \right\rangle$,  respectively. Thus, only the corrections with $\hat{H}_{1}$  contain the size of the nanoobject.

\section{Shift of levels by a magnetic field}
Using (\ref{9}), we consider the shift of stationary states caused by the magnetic field:
\begin{align}
\label{10}
\Delta E \left(n_{a} n_{b} n_{c}\right)&\equiv {\left\langle n_{a} n_{b} n_{c} \right|}\hat{H}{\left| n_{a} n_{b} n_{c}\right\rangle}-\Big\langle n_{a} n_{b} n_{c}\Big|\hat{H}_{0}+\frac{e\hbar}{2m}\vec{B}\vec{\sigma }\Big| n_{a} n_{b} n_{c}\Big\rangle ={\left\langle n_{a} n_{b} n_{c}\right|} \hat{H}_{1}{\left| n_{a} n_{b} n_{c} \right\rangle}
\nonumber\\
&+\sideset{}{’}\sum_{n_{x} n_{y} n_{z}} \frac{{\left\langle n_{a} n_{b} n_{c}\right|} \hat{H}_{1}{| n_{x} n_{y} n_{z}\rangle}{\langle n_{x} n_{y} n_{z}|} \hat{H}_{1} {\left| n_{a} n_{b} n_{c}\right\rangle}}{E_{n_{a} n_{b} n_{c}}-E_{n_{x} n_{y} n_{z}}}.
\end{align}

Let us analyze two cases of the magnetic field $\vec{B}$ orientation:
$a)$ the Landau gauge case, $\vec{A}(\vec{r})=\left(0,Bx,0\right)$. Taking into account the relationship $\vec{B}=\rot\vec{A}\left(\vec{r}\right)\Rightarrow \left(0,0,B\right)$, the field is oriented along $OZ$ axis $b)$ the gauge $\vec{A}(\vec{r})=\left(Bz,0,\alpha By\right)$ --- magnetic field $\vec{B}$ is in $XOY$ plane with different orientation, depending on parameter $\alpha$, in particular, at $\alpha = 1$, $\vec{B}$ is oriented along the diagonal in  $XOY$ plane, and at $\alpha=0$ along $OY$ axis.

\subsection*{Case $a)$ $\vec{A}(\vec{r})=\left(0,Bx,0\right)$.}

The 1st correction, taking into account that $p_{y} A_{y}=A_{y} p_{y}$, is:
\begin{equation} \label{11}
{\left\langle n \right|} \hat{H}_{1} {\left| n \right\rangle} =\frac{e}{2m} {\left\langle n \right|} \hat{H}_{1} {\left| n \right\rangle} \Rightarrow \frac{e}{2m} {\left\langle n \right|} (A_{y} \hat{p}_{y} +\hat{p}_{y} A_{y} ){\left| n \right\rangle} \Rightarrow \frac{eB}{m} \frac{\hbar }{\ri} {\Big\langle n \Big|} x\frac{\partial }{\partial y} {\Big| n \Big\rangle} \Rightarrow \frac{eB}{m} \frac{\hbar }{\ri} i_{1} i_{2} i_{3}.
\end{equation}

In equation (\ref{11}),  we used the denotation ${\left\langle n \right|} ={\left\langle n_{a} n_{b} n_{c}  \right|}$; it is taken into account that the term ${\left\langle n \right|} \frac{e\hbar}{2m}\vec{B}\vec{\sigma }{\left| n \right\rangle}$ equals zero due to the absence of spin-orbital interaction; with the denotation $q=\frac{n_{a} \piup }{a} x$, using the integrals tables \cite{Pru92},
\begin{equation}
i_{1} =\frac{2}{a} \cdot \left(\frac{a}{n_{a} \piup} \right)^{2} \cdot \int _{0}^{n_{a} \piup }q\cdot \sin ^{2}  q\cdot \rd q\Rightarrow \frac{2}{a} \cdot \left(\frac{a}{n_{a} \piup } \right)^{2} \left. \left(\frac{q^{2}}{4}-\frac{q}{4} \sin 2q-\frac{1}{8} \cos 2q \right)\right|_{0}^{n_{a}\piup} =\frac{a}{2}\,,
\end{equation}
\begin{equation}
i_{2} =\frac{2}{b} \cdot \int _{0}^{b}\sin \frac{n_{b}\piup }{d}  y\frac{\partial }{\partial y} \sin \frac{n_{b} \piup }{d} y\cdot \rd y\Rightarrow \frac{2}{b} \cdot \frac{1}{2} \left.\left(\sin ^{2} \frac{n_{b} \piup }{b} y\right)\right|_{0}^{b} =0,
\end{equation}
\begin{equation}
i_{3} =\frac{2}{c} \cdot \int _{0}^{c}\sin^{2}\frac{n_{c}\piup}{c} z \cdot \rd z=1.
\end{equation}
Thus, the 1st correction (\ref{10}) is zero.

Let us consider the 2nd correction.

We start with the calculation of the matrix elements in equation~(\ref{10})
\begin{equation}
\label{12}
{\left\langle n \right|} \hat{H}_{1} {\left| m \right\rangle} =\frac{e}{2m}{\left\langle n \right|} \hat{H}_{1} {\left| n \right\rangle} \Rightarrow \frac{eB}{m} \frac{\hbar }{\ri} {\Big\langle n \Big|} x\frac{\partial }{\partial y} {\Big| m \Big\rangle} \Rightarrow \frac{eB}{m} \frac{\hbar }{\ri} i_{4} i_{5} i_{6}\,,
\end{equation}
where
\begin{align}
i_{4} &=\frac{2}{a} \cdot \int _{0}^{a}x\cdot \sin \frac{n_{1} \piup }{a}  x\cdot \sin \frac{m_{1} \piup }{a} x\cdot \rd x\Rightarrow \frac{2}{a} \cdot  x\frac{1}{2}  \left. \left[\cos \frac{\left(n_{1}-m_{1} \right)\piup }{a} x-\cos \frac{\left(n_{1} +m_{1} \right)\piup }{a} x\right]\right|_{0}^{a} \nonumber\\
 &\Rightarrow \frac{a}{\piup ^{2}} \frac{8n_{1} m_{1} }{(m_{1}^{2} -n_{1}^{2})^{2} } \left[\left(-1\right)^{n_{1} +m_{1}}-1\right],
\end{align}
\begin{equation}
i_{5} =\frac{2}{b} \cdot \int _{0}^{b}\sin \frac{n_{2} \piup }{b}  y\frac{\partial }{\partial y} \sin \frac{m_{2} \piup }{b} y\cdot \rd y\Rightarrow \left(-1\right)\frac{4}{b} \frac{n_{2} m_{2} }{m_{2}^{2} -n_{2}^{2} } \left[\left(-1\right)^{n_{2} +m_{2} } -1\right],
\end{equation}
\begin{equation}
i_{6}=\frac{2}{c} \cdot \int _{0}^{c}\sin\frac{n_{3} \piup}{c}  z\cdot \sin \frac{m_{3}\piup }{c} z\cdot \rd z=\Delta _{n_{3}m_{3}}
\end{equation}
($\Delta _{n_{3} m_{3}}$ is the Kronecker symbol).

Thus, the 2nd correction is non-zero only at even $n_{1}+m_{1}$, $n_{2}+m_{2}$, i.e., $n_{i}+m_{i} =2k_{i}+1$ $\left(k_{i} =1, 2, 3,\ldots\right)$.

In particular, the correction $\Delta _{2} E\left(111\right)$ to the ground state $E\left(111\right)$ in terms of the units  $[-(\frac{eB}{m})^{2}\cdot(\frac{32}{\piup ^{2}})^{2}\cdot\frac{2\cdot 10^{-18}\cdot m}{\piup ^{2}}]$ is of the following form:
\begin{equation} \label{13}
\Delta _{2} E\left(111\right)=\left(\frac{a}{b} \right)\cdot  \sideset{}{’}\sum_{k_{1} k_{2}} \left\{\frac{2k_{1} }{\left[\left(2k_{1} \right)^{2} -1\right]^{2} } \cdot \frac{2k_{2} }{\left[\left(2k_{2} \right)^{2} -1\right]} \right\}^{2}   \left[\frac{\left(2k_{1} \right)^{2}-1}{a^{2}} +\frac{\left(2k_{2} \right)^{2} -1}{b^{2}} \right]^{-1}.
\end{equation}

\begin{figure}[!t]
\center {\includegraphics[scale=0.55] {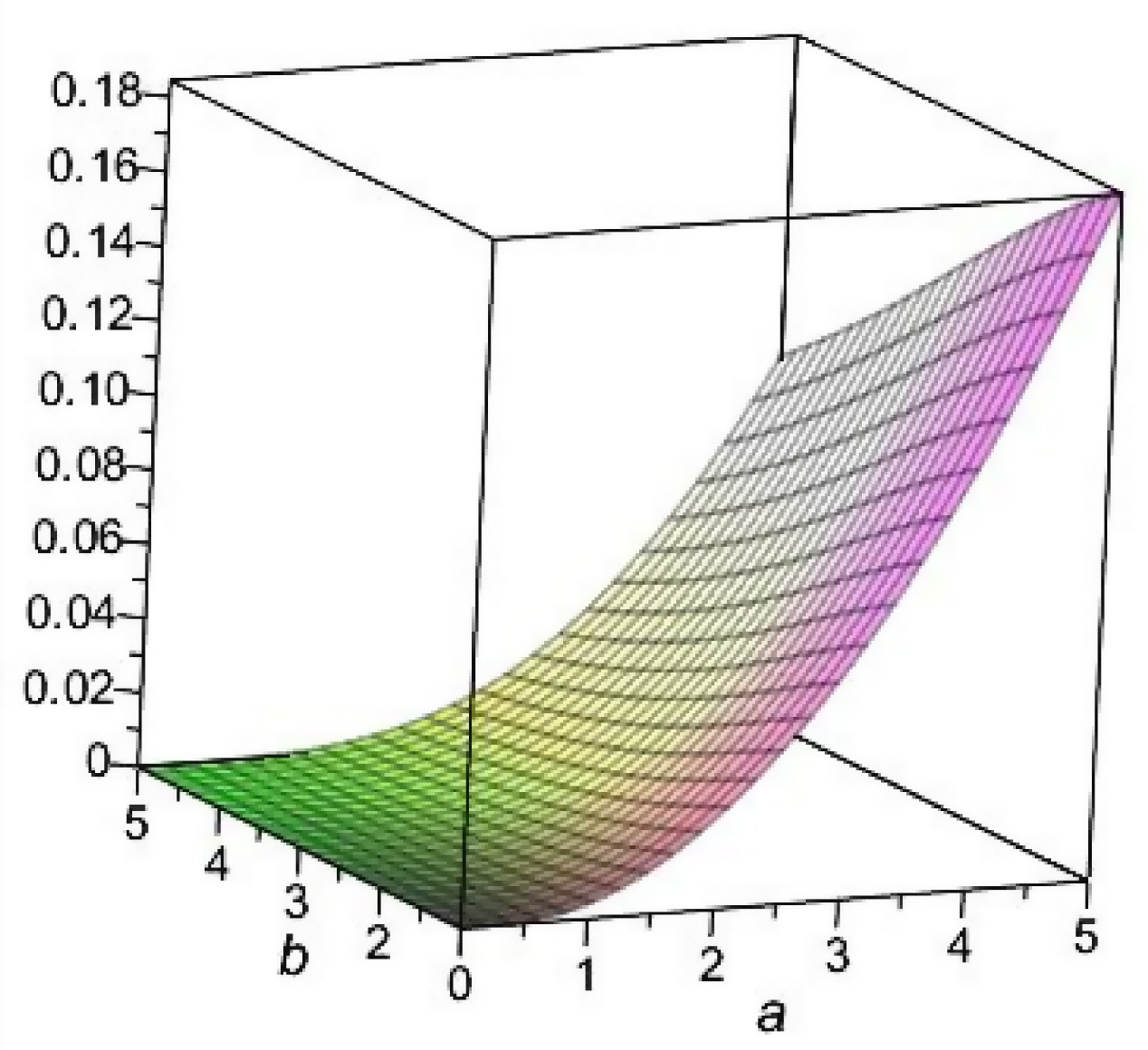}}
\caption{(Colour online) Dependence of the 2nd correction in terms of the units $[-(\frac{eB}{m} )^{2} \cdot(\frac{32}{\piup ^{2} } )^{2} \cdot\frac{2\cdot 10^{-18} \cdot m}{\piup ^{2} } ]$ on the sizes in $XOY$ plane in the Landau gauge case, $\vec{A}=\left(0,Bx,0\right)$ (magnetic field oriented along the $OZ$) ($a,b$ in nm).}
\label{fig1}
\end{figure}

It is seen that such a correction does not depend on the length of $c$ of the nanoobject, along which the magnetic field $\vec{B}$ is directed. Figure~\ref{fig1} shows the dependence of the correction on the size of the nanoobject in $XOY$ plane, which is perpendicular to the direction of $\vec{B}$. The correction monotonously increases with an increase of $a$ for any value of $b$ from the interval 1--5~nm, and vice versa, monotonously decreases with an increase of $b$ for any value  of $a$ from the interval $1{-}5$~nm.

We will evaluate the correction to the energy state $E(111)$ of the nanoobject for $B= 10$~T and  for different values of $a=b$. Table~\ref{tbl-smp1} shows the relative magnitudes of the correction $\big(\frac{\Delta _{2} E\left(111\right)}{E\left(111\right)}\big)_{a} $.

\begin{table}[!t]
\caption{Relative values of the correction, $\Big(\frac{\Delta _{2} E\left(111\right)}{E\left(111\right)} \Big)_{a} $.}
\label{tbl-smp1}
\vspace{2ex}
\centering
\begin{tabular}{|c|c|c|c|c|}
\hline\hline
$a$, nm & 1 & 10 & 20 & 30 \\ \hline
$\frac{\Delta _{2} E\left(111\right)}{E\left(111\right)}$, \% & $10^{-4\strut} $ & 0.1 & 1.5 & 10 \\ \hline\hline
\end{tabular}
\end{table}

The obtained results indicate a sharp dependence of the spectrum in a magnetic field on the nanoobject size.

\subsection*{ Case $b)$ $\vec{A}(\vec{r})=\left(Bz,0,\alpha By\right)$.}

In this case, the perturbation Hamiltonian is as follows:
\begin{equation} \label{14}
H_{1} =\frac{e}{2m}\big(\vec{A}\vec{p}+\vec{p}\vec{A}\big)\Rightarrow \frac{e}{m}\left(A_{x} p_{x} +A_{z} p_{z}\right)\Rightarrow \frac{eB}{m} \frac{\hbar}{\ri}\left(z\frac{\partial}{\partial x} +\alpha y\frac{\partial}{\partial z} \right).
\end{equation}

In the case  of $\alpha = 0$, i.e., with $\vec{B}$ directed along $OX$, the results and conclusions qualitatively coincide with the similar ones in the case $\vec{A}(\vec{r})=\left(0,Bx,0\right)$. Herein below we will consider the case of $\alpha = 1$.

Here, the 1st correction, similarly to the case $a)$, equals zero.

The calculations, similar to the above ones, in terms of the same units, give the following 2nd correction to the ground state:
\begin{align} 
\Delta'_{2} E\left(111\right)=\left(\frac{c}{a} \right)^{2} \cdot \sideset{}{’}\sum_{k_{1} k_{3} }\left\{\frac{2k_{1} }{\left[\left(2k_{1} \right)^{2} -1\right]} \cdot \frac{2k_{3}}{\left[\left(2k_{3} \right)^{2} -1\right]^{2} } \right\}^{2}   \left[\frac{\left(2k_{1} \right)^{2} -1}{a^{2} } +\frac{\left(2k_{3} \right)^{2} -1}{c^{2} } \right]^{-1} \nonumber
\end{align}
\begin{align}
+\alpha \cdot \left(\frac{b}{c} \right)^{2} \cdot \sideset{}{’}\sum_{k_{2} k_{3}}\left\{\frac{2k_{3} }{\left[\left(2k_{3} \right)^{2} -1\right]} \cdot \frac{2k_{2} }{\left[\left(2k_{2} \right)^{2} -1\right]^{2} } \right\}^{2}  \left[\frac{\left(2k_{2} \right)^{2} -1}{b^{2} } +\frac{\left(2k_{3} \right)^{2} -1}{c^{2} } \right]^{-1} .
\label{15}
\end{align}

\begin{figure}[!t]
\center {\includegraphics[scale=0.75]{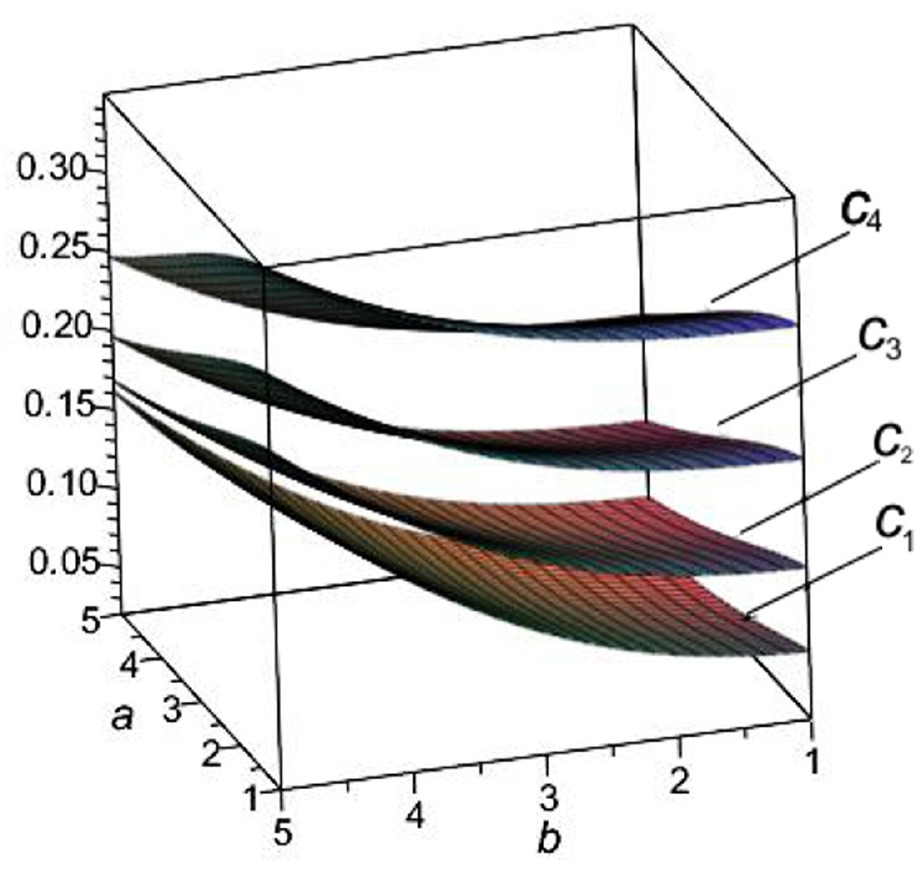}}
\caption{(Colour online) Dependence of the 2nd correction in terms of the units $[-(\frac{eB}{m} )^{2} \cdot (\frac{32}{\piup ^{2}} )^{2} \cdot\frac{2\cdot 10^{-18} \cdot m}{\piup ^{2} } ]$ on the sizes in $XOY$ plane in the gauge case, $\vec{A}=\left(Bz,0,By\right)$ (the direction of the magnetic field coincides with the bisection of the $XOY$),  $c_{1}=3$,  $c_{2}=4$,  $c_{3}=5$,  $c_{4}=6$ ($a$, $b$, $c$ in nm). }
\label{fig2}
\end{figure}

This correction, in contrast to the analogous one in the case \textit{a}), contains the dependence of the length of \textit{c}, as well as of its ratios to \textit{a} and \textit{b}, namely $(\frac{c}{a})$, $(\frac{b}{c})$. Figure~\ref{fig2} shows the analogous to figure~\ref{fig1} dependences of the 2nd correction for the size family $c$. It is shown that the greater $c$, the greater $\Delta'_{2} E\left(111\right)$. For all surfaces $\left. \Delta'_{2} E\left(111\right)\right|_{c}$, for any fixed value of $a$, the correction monotonously increases with an increase in $b$ and weakly decreases with an increase in $a$ at fixed $b$.

\begin{figure}[!b]
\begin{minipage}{0.49\textwidth}
\center{\includegraphics[width=.99\textwidth]{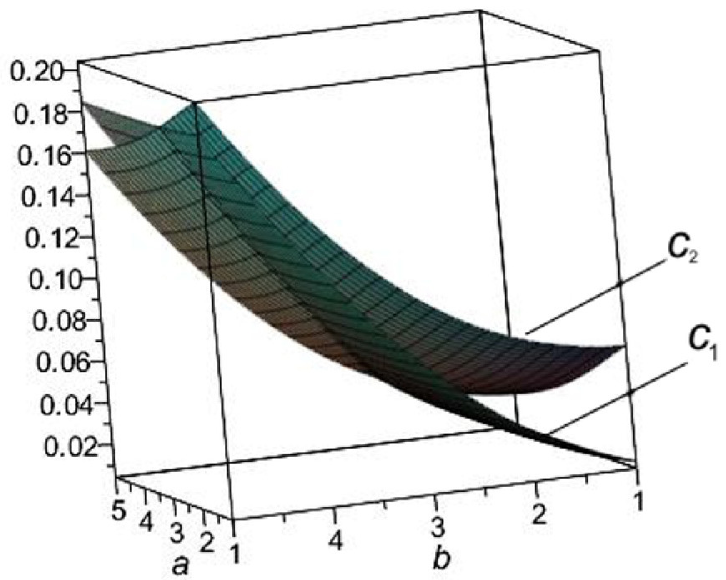}}
\end{minipage}
\hfill
\begin{minipage}{0.49\textwidth}
\center{\includegraphics[width=.99\textwidth]{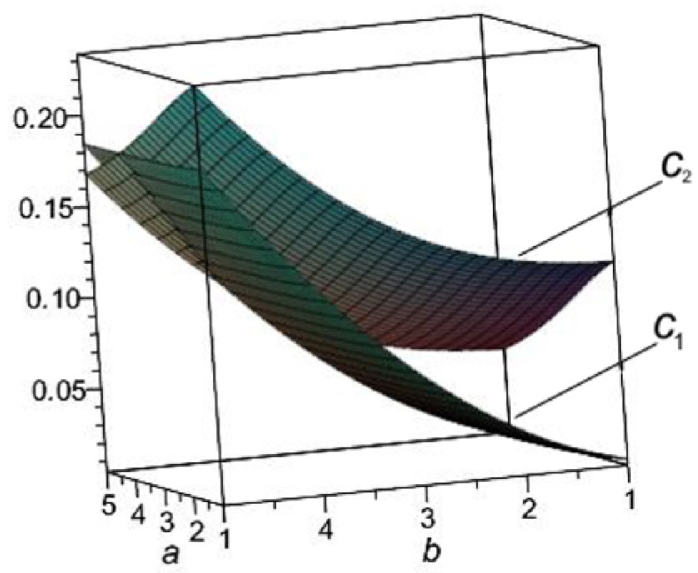}}
\end{minipage}
\begin{tabular}{p{0.48\textwidth}p{0.48\textwidth}}
\centering a) & \centering b) \\
\end{tabular}
\vspace{-0.5cm}
\caption{(Colour online) Dependence of the 2nd correction in terms of the units $[-(\frac{eB}{m})^{2} \cdot (\frac{32}{\piup ^{2} } )^{2} \cdot\frac{2\cdot 10^{-18} \cdot m}{\piup ^{2}} ]$  on the sizes in $XOY$ plane in the case $\vec{A}=\left(Bz,0,By\right)$ (the direction of the magnetic field coincides with the bisection of the $XOY$)  at $c_{1}=1$; $c_{2}=3$ (a) and $c_{1}=1$;  $c_{2}=4$ (b) ($a,b,c$ in nm).}
\label{fig3}
\end{figure}

Figure~\ref{fig3} shows the dependencies $\Delta'_{2} E\left(111\right)=f\left(a,b\right)$ for the family $c_{1}=1$ and $c_{2}=3$~(a) and $c_{1} =1$ and $c_{2} =4$~(b). There is clearly seen an intersection of two pairs of planes. The points $a_{i}$, $ b_{i} $ on the line of intersection are those in which the corrections corresponding to $c_{1} $, $c_{2}$ are of the same values. From the difference in the intersections in the cases $a$) and $b$), we can make a conclusion regarding the possible intersections outside the intervals $a\in \left[1,5\right]$ and $b\in \left[1,5\right]$.

\section{Conclusions}

The analysis of the electron spectrum of a nanoobject and of the shape of rectangular  parallelepiped placed in an external magnetic field, depending on the size of the object,  indicates that
\begin{enumerate}
\renewcommand{\labelenumi}{(\arabic{enumi})}
\item  in the framework of perturbation theory,  the correction to the spectrum appears only from the second term onward;
\item  the magnitudes of the corrections depend both on the magnitude of the magnetic field $\vec{B}$ (it is greater for greater fields) and on its orientation relative to the nanoobject;
\item in the case of Landau gauge  ($\vec{B}$ is directed along $OZ$ axis), the correction does not depend on the length of the nanoobject in this direction;
\item in the case $\vec{A}=B\left(z,0,\alpha  y\right)$, for any parameter $\alpha$ from the semi-interval $(0,1]$, unlike in the former case, the correction  depends on the three lengths of the nanoobject;
\item it is established that in the case of   $\vec{A}=B\left(z,0,y\right)$, there exists such a set of dimensions  of the  nanoobject  in $XOY$ plane for which the corrections are the same as those for their certain values of length along $OZ$ axis.
\end{enumerate}

Thus, in order to purposefully change the electronic spectrum of the  nanoobject by the magnetic field~$\vec{B}$, one should take into account not only  the orientation of  $\vec{B}$ in the nanoobject, but also its size and relationship between its geometrical  characteristics.

\ukrainianpart

\title{Розмірні ефекти в нанооб'єктах у магнітному полі}
\author{Б.А. Лукіянець, Д.В. Матулка}
\address{
Національний університет ``Львівська політехніка'', вул. С. Бандери, 12, 79000 Львів, Україна
}

\makeukrtitle

\begin{abstract}
\tolerance=3000%
Проведений теоретичний аналіз впливу розмірів на фізичні властивості нанооб'єкта у вигляді прямокутного паралелепіпеда зі сторонами $a$, $b$, $c$, орієнтованих, відповідно, вздовж осей координат $OX$, $OY$, $OZ$. В рамках теорії збурень аналізуються зміни в електронному спектрі нанооб'єкту, викликаного зовнішнім магнітним полем $\vec{B}$, залежно від розмірів нанооб'єкта. Розглянуто два випадки орієнтації магнітного поля: 1) з калібровкою Ландау, $\vec{A}(\vec{r})=\left(0,Bx,0\right)$, ($\vec{B}$ орієнтований уздовж сторони $c$) і 2) на $\vec{A}(\vec{r})=\left(Bz,0,\alpha By\right)$, $\alpha$ є параметр, що міняється в інтервалі $\left[0,1\right]$, зокрема при $ \alpha  = 0$ $\vec{B}$ направлений вздовж осі $OX$, а при $\alpha = 1$ $\vec{B}$ направлений вздовж бісектриси площини $XOY$. По-перше, показано, що перша поправка спектру дорівнює нулю, незалежно від орієнтації  $\vec{B}$. По-друге, встановлено, на відміну від випадку 1) орієнтації поля, де поправка не залежить від довжини $c$, у випадку 2) така поправка залежить як від $c$, так і від її співвідношень до довжини $a$ і $b$. Знайдено існування таких розмірів нанооб'єктів у площині $XOY$, при яких поправки до спектру однакові для різної довжини $c$ нанооб'єкта.
\keywords нанооб'єкт, магнітне поле, електронний спектр, розмірний ефект

\end{abstract}

\end{document}